\documentclass{elsarticle}
\usepackage{graphicx}%
\usepackage{multirow}%
\usepackage{amsmath,amssymb,amsfonts}%
\usepackage{amsthm}%
\usepackage{mathrsfs}%
\usepackage[title]{appendix}%
\usepackage{xcolor}%
\usepackage{textcomp}%
\usepackage{manyfoot}%
\usepackage{booktabs}%
\usepackage{algorithm}%
\usepackage{algorithmicx}%
\usepackage{algpseudocode}%
\usepackage{listings}%
\usepackage{verbatim}
\usepackage{enumitem}
 \usepackage{comment}
 \usepackage{amsmath}

\journal{}
\begin{document}

\begin{frontmatter}

\title{Exact Solution of Granovetter's Threshold Model for a Finite Population }

\author[mymainaddress]{Bianca Y.  S.  Ishikawa}

\author[mymainaddress]{Jos\'e F.  Fontanari}

\address[mymainaddress]{Instituto de F\'{\i}sica de S\~ao Carlos, Universidade de S\~ao Paulo, 13566-590 S\~ao Carlos, S\~ao Paulo, Brazil}

\begin{abstract}
Granovetter's threshold model provides a classical framework for social mobilization, in which collective action spreads through self-reinforcing cascades as individuals join once the movement size reaches their personal threshold. Here, we investigate social tipping points by treating the initial fraction of instigators, $\rho_0$, as a control parameter in a finite population of $N$ agents with activation thresholds drawn from a Beta distribution with shape parameters $\alpha$ and $\beta$. We characterize the thermodynamic phase diagram, which exhibits both continuous and discontinuous cascade transitions, and develop an exact combinatorial formulation for finite populations. The discontinuous transitions are hybrid, combining a jump in the final active fraction with a square-root critical singularity associated with an interior saddle-node bifurcation. Using the empirical cumulative distribution of quenched thresholds, we develop an asymptotic finite-size scaling theory that identifies the stochastic mechanisms responsible for rounding these transitions. For power-law thresholds ($\alpha>1$, $\beta=1$), the critical window and the mean inactive fraction at criticality both scale as $N^{-1/3}$. For hybrid transitions governed by an interior saddle-node, the critical window scales as $N^{-1/2}$, while the conditional displacement of stalled cascades from the marginal state scales as $N^{-1/4}$. We further derive an explicit Gaussian scaling function for the probability of complete activation and show that convergence to this limiting function is exceptionally slow, with corrections of order $N^{-1/6}$. These results establish the extensively seeded Granovetter model as a tractable framework for investigating finite-population fluctuations and hybrid criticality in irreversible threshold-driven cascades.
\end{abstract}

\end{frontmatter}


\section{Introduction}\label{sec:Intro}

The spontaneous emergence of collective behavior---ranging from public pro\-tests and opinion shifts to fashion fads and social norm changes---often unfolds through self-reinforcing chain reactions known as behavioral cascades~\cite{Bikhchandani_1992,Macy_2020}. A central challenge in modeling these phenomena is accounting for individual heterogeneity, since susceptibility to peer pressure varies substantially across a population. While classical stochastic models of group formation frequently assume homogeneous agents~\cite{Coleman_1961,Fontanari_2023} (see, however,~\cite{Starnini_2013,Mariano_2025}), Granovetter's seminal threshold model~\cite{Granovetter_1978,Granovetter_1983} introduced a framework explicitly grounded in microscopic diversity. In this model, an individual joins a collective movement only when the active fraction of the population reaches or exceeds their personal activation threshold, directly linking individual resistance to macroscopic social dynamics.

A fundamental question in the study of collective action concerns the tipping-point scenario~\cite{Oliver_1985,Watts_2002,Centola_2018}: what is the minimum initial fraction of activists required to overcome collective inertia and trigger a population-wide cascade? Recent work provided an exact analytical solution of Granovetter's model for a finite population of size $N$ in the single-instigator limit ($M_0=1$)~\cite{Fontanari_2026}. However, organized mobilization efforts need not rely on an isolated instigator and may instead begin with a substantial initial group of committed individuals. Extending the model to an extensive seed ($M_0=\rho_0N$), where the initial density $\rho_0$ remains fixed as $N\to\infty$, therefore provides a natural framework for investigating critical masses and collective tipping thresholds.

In this work, we investigate the irreversible, quenched Granovetter threshold model for finite populations initialized with an extensive seed density $\rho_0$. Individual activation thresholds are independently sampled from a Beta distribution with shape parameters $\alpha$ and $\beta$. The resulting deterministic phase diagram contains continuous transitions, controlled by marginal fixed points at the fully active boundary, and discontinuous hybrid transitions associated with interior saddle-node bifurcations. For the symmetric Beta distribution ($\alpha=\beta=2$), the discontinuous transition occurs at the critical seed density $\rho_0^c=1/9$, where the final active fraction undergoes a macroscopic jump of magnitude $3/4$.

Hybrid, or mixed-order, phase transitions combine an abrupt discontinuity in the order parameter with critical singularities characteristic of continuous transitions. Their study has a long history in statistical physics, beginning with Fisher's analysis of cooperative transitions in biopolymers~\cite{Fisher_1966} and Thouless's investigation of one-dimensional Ising models with inverse-square interactions~\cite{Thouless_1969,Bar_2014}. More recently, hybrid transitions have been identified in network processes such as $k$-core percolation and cascading failures in interdependent networks~\cite{Dorogovtsev_2006,Baxter_2010,Lee_2016}. Although these systems differ substantially in their microscopic mechanisms, several exhibit a common mean-field saddle-node structure, in which a stable branch of collective states disappears through its collision with an unstable branch. The resulting discontinuous jump coexists with a square-root singularity as the critical point is approached. Understanding how finite-size fluctuations modify this behavior is particularly important because the sharp transition of the thermodynamic limit is replaced, in finite systems, by a probabilistic competition between distinct macroscopic outcomes.

Our principal contribution is an analytical description of finite-population effects based on the empirical cumulative distribution of quenched thresholds. The exact finite-$N$ dynamics can be expressed directly in terms of this random distribution, whose fluctuations are governed by classical empirical-process theory \cite{van_der_Vaart_1998}. Combining these fluctuations with the local structure of the deterministic fixed-point equations, we derive distinct critical scaling laws. For power-law thresholds ($\alpha>1$, $\beta=1$), the continuous transition has a critical window of width $N^{-1/3}$, and the mean inactive fraction at criticality decays with the same exponent. For the interior saddle-node transition, the critical window instead scales as $N^{-1/2}$, while the displacement of the stalled state from its marginal value scales as $N^{-1/4}$. We also obtain an explicit Gaussian scaling function for the probability of a complete cascade and show that its convergence is exceptionally slow, with finite-size corrections of order $N^{-1/6}$. These results identify the stochastic mechanisms responsible for rounding continuous and hybrid cascade transitions and establish the extensively seeded Granovetter model as a tractable setting for studying critical fluctuations in irreversible collective dynamics.

\section{Model Formulation with Extensive Seeding}\label{sec:model_ext}

Consider a population of $N$ agents subject to an extensive initial seed density $\rho_0 \in (0, 1)$ of active instigators at time $t=0$. The population is partitioned into two distinct groups: a fraction $\rho_0$ of unconditionally active individuals (whose intrinsic threshold is $x = 0$) and a remaining fraction $1 - \rho_0$ of susceptible individuals.

The normalized activation thresholds $x \in [0, 1]$ of the susceptible agents are drawn independently from a continuous probability density function $f(x; \alpha, \beta)$. We parameterize this susceptibility using the flexible Beta distribution, 
\begin{equation}\label{eq:beta_pdf}
f(x; \alpha, \beta) = \frac{x^{\alpha-1}(1-x)^{\beta-1}}{\mathrm{B}(\alpha, \beta)},
\end{equation}
where $\mathrm{B}(\alpha, \beta)$ is the Beta function, and $\alpha, \beta > 0$ are shape parameters dictating microscopic diversity and polarization.

Combining both groups, the total effective threshold probability density $g(x)$ across the entire population is given by
\begin{equation}\label{eq:total_pdf}
g(x) = \rho_0 \delta(x) + (1-\rho_0) f(x; \alpha, \beta),
\end{equation}
where $\delta(x)$ is the Dirac delta function representing the unconditionally active instigators at the origin. Integrating equation~(\ref{eq:total_pdf}) yields the total effective cumulative distribution function (CDF), $G(x)$, representing the proportion of agents with thresholds less than or equal to $x$,
\begin{equation}\label{eq:total_cdf}
G(x) = \rho_0 \Theta(x) + (1-\rho_0) F(x; \alpha, \beta),
\end{equation}
where $\Theta(x)$ is the Heaviside step function and $F(x; \alpha, \beta)$ is the regularized incomplete Beta function \cite{Abramowitz_1972}.

The cascade dynamics unfold in discrete time steps $t = 0, 1, 2, \dots$ through synchronous, parallel updating~\cite{Granovetter_1978}. At each time step $t$, every inactive agent $i$ with threshold $x_i$ becomes active at time $t+1$ if and only if their threshold is less than or equal to the current total fraction of active individuals, $x_i \le \rho_t$. Activation is irreversible, meaning active agents remain active indefinitely. Once the set of personal thresholds $\{x_i\}_{i=1}^N$ is assigned at $t=0$, the progression of the behavioral cascade is entirely deterministic. Starting from the initial active seed fraction $\rho_0$, the system evolves until reaching a steady state at time step $t^* \le N - M_0$ where no additional agents meet their activation condition ($\rho_{t^*+1} = \rho_{t^*}$), establishing the final asymptotic active fraction $\rho_\infty = \rho_{t^*}$.

\section{The Deterministic Regime}
\label{sec:det_reg}

In the thermodynamic limit ($N\to\infty$), the empirical distribution of quenched thresholds converges to its population distribution, and the fraction of active agents obeys the deterministic Granovetter map~\cite{Granovetter_1978,Granovetter_1983},
\begin{equation}
\label{eq:dyn_map_G}
\rho_{t+1}=G(\rho_t).
\end{equation}
Using equation~(\ref{eq:total_cdf}), this becomes
\begin{equation}
\label{eq:dyn_map_explicit}
\rho_{t+1} = \rho_0+(1-\rho_0)F(\rho_t;\alpha,\beta).
\end{equation}
The stationary active fraction satisfies
\begin{equation}
\label{eq:fixed_point}
\rho_\infty = \rho_0+(1-\rho_0)F(\rho_\infty;\alpha,\beta).
\end{equation}
Since the map is nondecreasing and the dynamics starts from $\rho_0$, the physical final state is the smallest fixed point not below $\rho_0$. In particular, $\rho_\infty=1$ is always a fixed point, although it need not be dynamically accessible from the prescribed seed.

An interior fixed point is linearly stable when $(1-\rho_0)f(\rho_\infty;\alpha,\beta)<1$~\cite{Strogatz_2015}. The marginal stability condition is
\begin{equation}
\label{eq:stability_boundary}
(1-\rho_0)f(\rho_\infty;\alpha,\beta)=1.
\end{equation}
Together with equation~(\ref{eq:fixed_point}), this condition determines the critical seed densities and bifurcation coordinates. We summarize below the regimes relevant to the finite-population analysis.

\subsection{The Power-Law Threshold Distribution
$(\beta=1)$}
\label{sec:det_power_law}

For $\beta=1$, the cumulative threshold distribution reduces to $F(\rho;\alpha,1)=\rho^\alpha$, and the fixed-point equation becomes
\begin{equation}
\label{eq:fixed_point_power}
\rho_\infty
=
\rho_0+(1-\rho_0)\rho_\infty^\alpha.
\end{equation}
For $\alpha\leq1$ and $\rho_0>0$, the dynamics reaches complete activation. For $\alpha>1$, a stable partial-cascade solution exists below the critical seed density
\begin{equation}
\label{eq:rho_0_c}
\rho_0^c=1-\frac{1}{\alpha},
\end{equation}
obtained by applying the marginal stability condition at $\rho_\infty=1$.

The partial-cascade branch joins the fully active state continuously at $\rho_0^c$. Writing $\Delta=1-\rho_\infty$ and expanding equation~(\ref{eq:fixed_point_power}) for $\Delta\ll1$ gives
\begin{equation}
\label{eq:transcritical_expansion}
1\simeq
(1-\rho_0)\alpha
-
(1-\rho_0)\frac{\alpha(\alpha-1)}{2}\Delta.
\end{equation}
Consequently,
\begin{equation}
\label{eq:scaling_result}
\Delta
\simeq
\frac{2\alpha}{\alpha-1}
(\rho_0^c-\rho_0),
\qquad
\rho_0\to\rho_0^{c-}.
\end{equation}
Thus, the transition is continuous, with order-parameter exponent $1$, and corresponds to a boundary transcritical bifurcation~\cite{Strogatz_2015}. Its finite-size scaling is examined in Section~\ref{sec:finite_N}.

\subsection{Smooth Crossover for $\beta<1$}
\label{sec:conservative}

For $\beta<1$, the threshold density diverges as $\rho\to1^-$, rendering the fully active fixed point repelling from below for every $\rho_0<1$. Consequently, an extensive seed $0<\rho_0<1$ produces a unique stable partial-cascade state, $\rho_0<\rho_\infty<1$.

This contrasts with the unseeded model, which exhibits an infinite-order transition at $\alpha=1$ for $\beta<1$. Approaching this transition from $\alpha<1$, the nonzero stationary solution vanishes through the essential singularity~\cite{Fontanari_2026}
\begin{equation}
\label{eq:essential_singularity}
\rho_\infty
\approx
\exp\left(
\frac{\ln\beta}{1-\alpha}
\right).
\end{equation}
Any positive extensive seed removes the zero-activation fixed point and replaces this singularity with a smooth crossover between partial-cascade states.

\subsection{Discontinuous Transitions for $\beta>1$}
\label{sec:susceptible}

For $\beta>1$, the threshold density vanishes at complete activation, $f(1;\alpha,\beta)=0$, so that $\rho_\infty=1$ is locally stable. In the parameter region where an interior saddle-node exists, notably for $\alpha>1$ and sufficiently small seed density, the map also has a stable low-activation fixed point $\rho_-$ and an unstable intermediate fixed point $\rho_+$. Although the coexistence of $\rho_-$ and $\rho_\infty=1$ corresponds to bistability of the map, the cascade dynamics has a prescribed initial condition, $\rho(0)=\rho_0$, rather than an arbitrarily selectable one. Since every fixed point satisfies $\rho=\rho_0+(1-\rho_0)F(\rho)\geq\rho_0$, the two interior fixed points necessarily obey $\rho_0<\rho_-<\rho_+<1$. The monotone dynamics therefore converges to $\rho_-$, the smallest fixed point above the initial seed density.

As $\rho_0$ increases, the stable low-activation fixed point $\rho_-$ and the unstable fixed point $\rho_+$ approach one another and eventually merge at a saddle-node bifurcation. The critical coordinates $(\rho^*,\rho_0^c)$ satisfy
\begin{align}
\rho^*
&=
\rho_0^c+
(1-\rho_0^c)F(\rho^*;\alpha,\beta),
\label{eq:bif_cond1}\\
1
&=
(1-\rho_0^c)f(\rho^*;\alpha,\beta).
\label{eq:bif_cond2}
\end{align}
Eliminating $\rho_0^c$ gives
\begin{equation}
\label{eq:bif_coordinate}
1-\rho^*
=
\frac{1-F(\rho^*;\alpha,\beta)}
{f(\rho^*;\alpha,\beta)},
\end{equation}
and the corresponding critical seed density is
\begin{equation}
\label{eq:critical_seed_beta_gt_1}
\rho_0^c
=
1-\frac{1}{f(\rho^*;\alpha,\beta)}.
\end{equation}

For $\rho_0<\rho_0^c$, the cascade converges to the stable low-activation fixed point $\rho_-$. At $\rho_0=\rho_0^c$, the two interior fixed points coalesce at the marginal state $\rho^*$, which is still reached asymptotically by the deterministic dynamics starting from $\rho_0^c$. For $\rho_0>\rho_0^c$, the interior fixed points disappear, leaving complete activation as the only accessible stationary state. Consequently, the final active fraction undergoes a discontinuous transition as the seed density crosses $\rho_0^c$, with a jump of magnitude $1-\rho^*$.

An analytically convenient example is the symmetric Beta distribution with $\alpha=\beta=2$, for which
\begin{equation}
F(\rho;2,2)=3\rho^2-2\rho^3.
\end{equation}
The critical coordinates are $\rho^*=1/4$ and $\rho_0^c=1/9$. Besides the fully active solution, the fixed-point equation admits the two branches
\begin{equation}
\label{eq:sol_alpha2_beta2}
\rho_{\pm}
=
\frac14
\left(
1\pm
\sqrt{\frac{1-9\rho_0}{1-\rho_0}}
\right),
\end{equation}
where $\rho_-$ is stable and $\rho_+$ is unstable. The stable branch approaches the saddle-node according to
\begin{equation}
\rho^*-\rho_-
\simeq
\frac{9}{8\sqrt{2}}
(\rho_0^c-\rho_0)^{1/2}.
\end{equation}
More generally, a quadratic expansion around a nondegenerate interior saddle-node gives
\begin{equation}
\label{eq:saddle_node_scaling}
\rho^*-\rho_-
\simeq
\left[
\frac{
2f(\rho^*)[1-F(\rho^*)]
}{
f'(\rho^*)
}
\right]^{1/2}
(\rho_0^c-\rho_0)^{1/2},
\end{equation}
provided $f'(\rho^*)>0$. The square-root singularity is characteristic of a saddle-node bifurcation and provides the deterministic critical exponent $\beta_m=1/2$ underlying the hybrid transition discussed below.

\subsection{Phase Diagrams}
\label{sec:phase_diagrams}

\begin{figure*}[t]
\centering
\includegraphics[width=\textwidth]{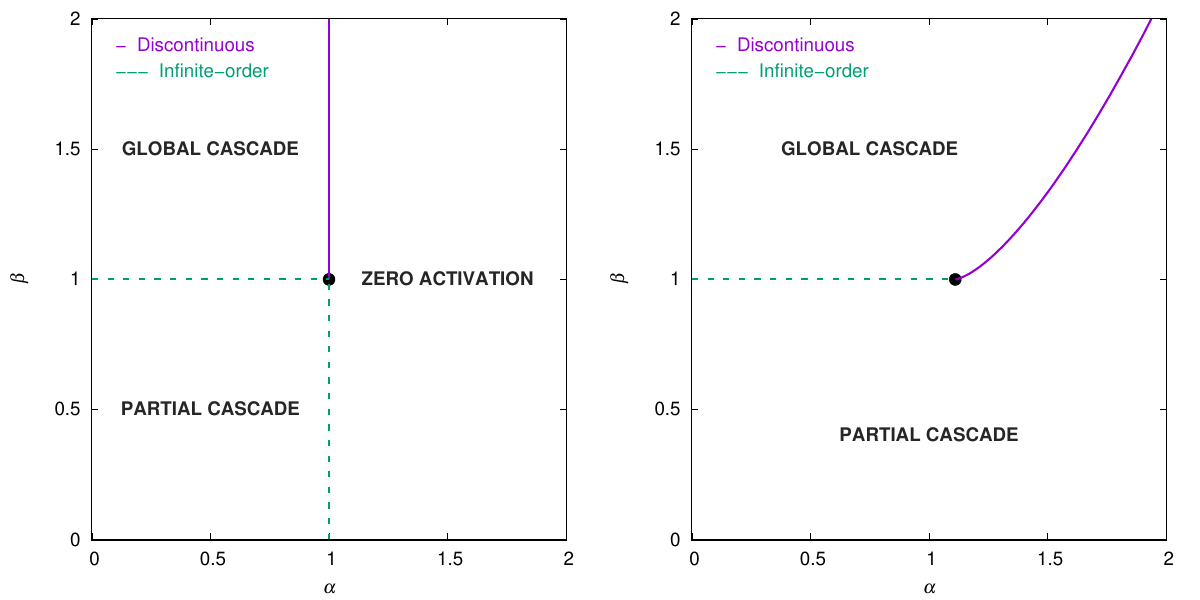}
\caption{
Deterministic phase diagrams in the $(\alpha,\beta)$ plane. Left panel: Unseeded model ($\rho_0=0$), showing zero-activation, partial-cascade, and global-cascade regimes. The solid boundary denotes a discontinuous transition, while the dashed boundaries correspond to infinite-order transitions. Right panel: Extensively seeded model ($\rho_0=0.1$), in which the zero-activation regime is absent. The solid curve marks the discontinuous saddle-node transition between partial and global cascades. The dashed line at $\beta=1$ marks the continuous boundary separating these regimes for $\alpha<10/9$. The two boundaries meet at $(\alpha,\beta)=(10/9,1)$.
}
\label{fig:phase_diagrams}
\end{figure*}

Figure~\ref{fig:phase_diagrams} summarizes the deterministic phase structure in the $(\alpha,\beta)$ plane. In the unseeded case (left panel), the zero-activation, partial-cascade, and global-cascade regimes meet at $(\alpha,\beta)=(1,1)$. Their boundaries include a discontinuous transition and continuous transitions characterized by essential singularities~\cite{Fontanari_2026}.

An extensive seed eliminates the zero-activation phase. As illustrated for $\rho_0=0.1$ in the right panel, the resulting phase diagram contains a discontinuous saddle-node boundary separating partial and global cascades. This boundary meets the line $\beta=1$ at
\begin{equation}
\alpha_c=\frac{1}{1-\rho_0},
\qquad
\beta_c=1.
\end{equation}
For $\rho_0=0.1$, the meeting point is $(\alpha_c,\beta_c)=(10/9,1)$. The distinction between boundary marginality at $\beta=1$ and an interior saddle-node for $\beta>1$ is central to the different finite-size scaling laws derived in the following section.

\section{The stochastic (finite-population) regime}
\label{sec:finite_N}

The deterministic self-consistency framework developed in Section~\ref{sec:det_reg} describes the thermodynamic limit $N\to\infty$. In finite populations, however, the realized
threshold distribution fluctuates from sample to sample. These fluctuations affect not only the final cascade size but also whether the dynamics can overcome an intermediate
activation bottleneck and reach full mobilization.

We develop two complementary descriptions of the finite-population dynamics. The first is an exact combinatorial formulation that determines the probability distribution of the final cascade size for arbitrary $N$ and initial seed size $M_0$. The second expresses the dynamics in terms of the empirical threshold distribution. Although both descriptions are exact at finite $N$, the latter is particularly convenient for analyzing finite-size fluctuations and critical scaling.

Throughout this section, the $M_0$ initial seeds are permanently active, while the remaining $N-M_0$ agents have independent, quenched thresholds drawn from the
cumulative distribution function $F(x)$. The seed set is chosen independently of these thresholds. We denote the actual initial active fraction by $\rho_{0,N}=M_0/N$.

\subsection{Exact combinatorial formulation}
\label{subsec:exact_combinatorial}

Let $P_N(k;M_0)$ denote the probability that a cascade initiated by $M_0\geq1$ seeds terminates with exactly $k$ active agents, where $M_0\leq k\leq N$.

For the cascade to terminate at size $k$, two conditions must hold. First, a specified subset of $k-M_0$ non-seed agents must be recruited through a sequence of successful activations starting from the $M_0$ initial seeds. Second, all remaining $N-k$ agents must have thresholds exceeding $k/N$, preventing further recruitment.

We denote by $Q_k^{(M_0)}$ the probability that the specified group of $k-M_0$ non-seed agents can be fully activated when the dynamics is restricted to this group and the $M_0$ seeds. The activation criterion is evaluated using the total population size $N$, so the active fraction at an intermediate cascade size $j$ is $j/N$.

Because successful propagation requires all participating non-seed agents to have thresholds at most $k/N$, we have
\begin{equation}
0 \leq  Q_k^{(M_0)} \leq [F(k/N)]^{k-M_0}.
\end{equation}
The upper bound is generally not attained because activation may fail at an intermediate cascade size even when all thresholds are below $k/N$.

There are $\binom{N-M_0}{k-M_0}$ possible choices of the recruited non-seed agents. Since the thresholds of these agents are independent of those of the remaining agents, the exact
probability distribution is
\begin{equation}
\label{eq:PN_gen}
P_N(k;M_0) = \binom{N-M_0}{k-M_0} Q_k^{(M_0)} \left[1-F\left(\frac{k}{N}\right)\right]^{N-k}.
\end{equation}

The propagation factor can be calculated recursively. Consider the dynamics restricted to a specified group containing the $M_0$ seeds and $k-M_0$ non-seed agents, with all thresholds drawn independently from the original distribution $F$. The final active population must contain some $j\in\{M_0,\ldots,k\}$ agents. Summing the probabilities of these mutually exclusive outcomes gives
\begin{equation}
\label{eq:conservation_sum}
\sum_{j=M_0}^{k} \binom{k-M_0}{j-M_0} Q_j^{(M_0)} \left[1-F\left(\frac{j}{N}\right)\right]^{k-j} =1.
\end{equation}
Here the probabilities are unconditional; in particular, the thresholds of the $k-M_0$ non-seed agents are not conditioned to lie below $k/N$.

Isolating the term $j=k$ yields
\begin{equation}
\label{eq:Qk_gen}
Q_k^{(M_0)} = 1- \sum_{j=M_0}^{k-1} \binom{k-M_0}{j-M_0} Q_j^{(M_0)} \left[1-F\left(\frac{j}{N}\right)\right]^{k-j},
\end{equation}
with the initial condition $Q_{M_0}^{(M_0)}=1$.

For $M_0=1$, this recursion reduces to the single-instigator formulation of Ref.~\cite{Fontanari_2026}. For uniformly distributed thresholds, $F(x)=x$, the recursion admits the closed-form solution
\begin{equation}
\label{eq:Qk_uniform_exact}
Q_k^{(M_0)} = \frac{M_0 k^{k-M_0-1}}{N^{k-M_0}}, \qquad k\geq M_0.
\end{equation}
This expression follows from Abel's binomial identity~\cite{Riordan_1968}. Its numerator counts labeled rooted forests with $M_0$ specified roots among $k$ vertices, providing a connection with Cayley's enumeration of labeled trees and forests~\cite{Cayley_1889}.

Equations~(\ref{eq:PN_gen}) and (\ref{eq:Qk_gen}) constitute an exact finite-population solution. In particular, the mean final active fraction is
\begin{equation}
\label{eq:stoc_expectation}
\langle\rho_\infty(N)\rangle = \frac{1}{N}\sum_{k=M_0}^{N} kP_N(k;M_0).
\end{equation}

Although the recursion provides an exact numerical benchmark, extracting critical finite-size scaling directly from it is difficult. The bound on $Q_k^{(M_0)}$ cannot, in general, be replaced by an asymptotic equality: the requirement of uninterrupted propagation introduces correlations among the successive activation events.
We therefore turn to an equivalent empirical-distribution representation, which makes the relevant stochastic fluctuations explicit.


\subsection{Empirical threshold dynamics}
\label{subsec:empirical_dynamics}

Let $n=N-M_0$ denote the number of non-seed agents  with quenched thresholds $X_1,\ldots,X_n$ drawn independently from $F$. Their empirical cumulative distribution is
\begin{equation}
\label{eq:empirical_cdf}
F_n(x) = \frac{1}{n} \sum_{i=1}^{n} \Theta(x-X_i).
\end{equation}
For a given realization of the thresholds, the active fraction evolves according to
\begin{equation}
\label{eq:empirical_map}
\rho_{t+1} = \rho_{0,N} + (1-\rho_{0,N})F_n(\rho_t), 
\end{equation}
which is exact at finite $N$. Because the thresholds are quenched and activation is irreversible, the sequence $\rho_t$ is nondecreasing and terminates after finitely many activation steps. Its final value is the smallest fixed point of the empirical map that is accessible from the initial seed density.

It is convenient to introduce the random gap function
\begin{equation}
\label{eq:empirical_gap}
H_N(\rho) = \rho-\rho_{0,N} -(1-\rho_{0,N})F_n(\rho),
\end{equation}
so that the final active fraction satisfies $H_N(\rho_\infty)=0$, with the physically realized solution selected by the monotone dynamics starting from $\rho_{0,N}$.
In the thermodynamic limit, the Glivenko--Cantelli theorem implies $F_n(\rho)\to F(\rho)$ uniformly almost surely~\cite{van_der_Vaart_1998}. The corresponding deterministic gap function is
\begin{equation}
\label{eq:deterministic_gap}
H(\rho;\rho_0) = \rho-\rho_0-(1-\rho_0)F(\rho),
\end{equation}
and its zeros satisfy
\begin{equation}
\label{eq:stochastic_fixed_point}
\rho^* = \rho_0+(1-\rho_0)F(\rho^*),
\end{equation}
recovering the self-consistency equation of Section~\ref{sec:det_reg}.

The finite-size fluctuations around this deterministic limit are described by the empirical-process central limit theorem (Donsker’s theorem)~\cite{van_der_Vaart_1998}.
For continuous $F$, and with $\rho_{0,N}\to\rho_0$, one has
\begin{equation}
\label{eq:empirical_bridge}
\sqrt{n}\,[F_n(\rho)-F(\rho)] \Rightarrow B(F(\rho)),
\end{equation}
where $B(s)$ is a standard Brownian bridge, with covariance
\begin{equation}
\label{eq:bridge_covariance}
\operatorname{Cov}[B(s),B(t)] = \min(s,t)-st.
\end{equation}
Consequently, the random gap can be written asymptotically as
\begin{equation}
\label{eq:stochastic_gap_expansion}
H_N(\rho) = H(\rho;\rho_{0,N}) - \sqrt{\frac{1-\rho_{0,N}}{N}}\, B(F(\rho)) + o_p(N^{-1/2}),
\end{equation}
where the expansion is understood in the empirical-process sense~\cite{van_der_Vaart_1998}. This representation identifies the source of finite-population fluctuations: the deterministic gap is perturbed by a correlated Gaussian process whose amplitude decreases as $N^{-1/2}$.

The behavior of the final cascade depends on the local structure of the deterministic gap near the dynamically selected fixed point. Regular fixed points and marginal fixed points generally exhibit different finite-size scaling laws.

\subsection{Regular fluctuations and critical
finite-size scaling}
\label{subsec:finite_size_general}

Consider first a regular, attracting fixed point $\rho^*\in(0,1)$ of the deterministic dynamics, satisfying
\begin{equation}
H(\rho^*;\rho_0)=0, \qquad H'(\rho^*;\rho_0)>0.
\end{equation}
Equivalently,
\begin{equation}
1-(1-\rho_0)f(\rho^*)>0,
\end{equation}
where $f(\rho)=F'(\rho)$.

Assume that this fixed point is selected by the deterministic dynamics and that the finite-population dynamics remains in its basin of attraction with probability tending to one.
 Expanding the empirical gap around $\rho^*$ gives
\begin{equation}
0 \simeq {} \left[1-(1-\rho_0)f(\rho^*)\right] (\rho_\infty-\rho^*) - \sqrt{\frac{1-\rho_0}{N}}\, B(F(\rho^*)).
\end{equation}
Therefore,
\begin{equation}
\label{eq:regular_fluctuations}
\sqrt{N}(\rho_\infty-\rho^*) \Rightarrow \frac{\sqrt{1-\rho_0}\, B(F(\rho^*))} {1-(1-\rho_0)f(\rho^*)},
\end{equation}
provided the difference between the actual seed fraction $\rho_{0,N}$ and its limiting value $\rho_0$ is $o(N^{-1/2})$.

Since the Brownian bridge satisfies $\operatorname{Var}[B(s)]=s(1-s)$, the asymptotic variance is
\begin{equation}
\label{eq:regular_variance}
\operatorname{Var}(\rho_\infty) \sim \frac{ (1-\rho_0)F(\rho^*) [1-F(\rho^*)]}{ N[1-(1-\rho_0)f(\rho^*)]^2}.
\end{equation}
Thus, a regular attracting fixed point exhibits Gaussian fluctuations of order $N^{-1/2}$, and the final active fraction is self-averaging within its basin of attraction.

This conclusion does not extend directly to marginal fixed points, which occur at a critical seed density $\rho_0^c$ and state $\rho^*$ satisfying
\begin{equation}
\label{eq:marginal_condition}
H(\rho^*;\rho_0^c)=0,
\qquad
H'(\rho^*;\rho_0^c)=0.
\end{equation}
At such points, the linear restoring term vanishes, and the scaling behavior in the critical region must be determined by expanding the deterministic gap jointly around $(\rho^*,\rho_0^c)$ and balancing it against the finite-size empirical fluctuations in equation~\eqref{eq:stochastic_gap_expansion}.

For an interior saddle-node bifurcation with $H''(\rho^*;\rho_0^c)\neq 0$, expanding $H(\rho;\rho_0)$ to leading order in both $\rho-\rho^*$ and $\rho_0-\rho_0^c$ yields
\begin{equation}
\label{eq:quadratic_gap}
H(\rho;\rho_0)
\simeq
\frac{\partial H}{\partial \rho_0}(\rho^*;\rho_0^c)(\rho_0-\rho_0^c)
+
\frac{1}{2} H''(\rho^*;\rho_0^c)(\rho-\rho^*)^2,
\end{equation}
where $\partial H/\partial \rho_0 = -[1-F(\rho^*)]<0$. At a halting state $\rho_\infty$, the condition $H_N(\rho_\infty)=0$ in equation~\eqref{eq:stochastic_gap_expansion} equates this deterministic gap to the stochastic perturbation of order $N^{-1/2}$. Far from criticality, where $|H(\rho;\rho_0)|\gg N^{-1/2}$, deterministic drift dominates over sample noise; conversely, the critical finite-size regime is defined by the dominant-balance condition $|H(\rho;\rho_0)|\sim N^{-1/2}$, where deterministic forces and empirical fluctuations compete on equal footing.

At exact criticality ($\rho_0=\rho_0^c$, assuming $\rho_{0,N}-\rho_0^c=o(N^{-1/2})$), the linear detuning term in equation~\eqref{eq:quadratic_gap} vanishes, leaving $H(\rho;\rho_0^c)\simeq \frac{1}{2}H''(\rho^*;\rho_0^c)(\rho-\rho^*)^2$. Balancing this quadratic gap against the $\mathcal{O}(N^{-1/2})$ empirical noise yields $(\rho-\rho^*)^2\sim N^{-1/2}$, which sets a characteristic displacement scale of order $N^{-1/4}$ for trajectories that stall near the marginal fixed point. When $\rho_0$ is varied around $\rho_0^c$, the first term in equation~\eqref{eq:quadratic_gap} shifts the deterministic gap linearly in $\rho_0-\rho_0^c$. Matching this detuning term against the $\mathcal{O}(N^{-1/2})$ empirical noise shows that the critical scaling window over which fluctuations smooth the transition has width $|\rho_0-\rho_0^c|\sim N^{-1/2}$, provided the empirical noise at the saddle-node has nonzero variance.

Within this $\mathcal O(N^{-1/2})$ window, sample-to-sample fluctuations of the empirical threshold distribution can either trap a realization near the marginal bottleneck $\rho^*$ or allow it to pass through the bottleneck toward a distant high-activity state. The limiting distribution of $\rho_\infty$ can therefore remain bimodal, and the final active fraction is no longer self-averaging. At leading order, the probability of crossing the bottleneck is governed by the Gaussian fluctuation of the empirical gap evaluated at $\rho^*$. Subleading corrections arise from spatial variations of the empirical process near the bottleneck. These variations become relevant on a finer $\mathcal O(N^{-1/3})$ scale, distinct from the $\mathcal O(N^{-1/4})$ displacement characteristic of trajectories that stall near the marginal fixed point.

A qualitatively different balance arises when the marginal fixed point coincides with a boundary of the domain, such as $\rho^*=1$. At the upper boundary, $F(1)=1$ and the Brownian-bridge variance $\operatorname{Var}[B(1)]=0$ vanishes, suppressing the leading $\mathcal O(N^{-1/2})$ Gaussian fluctuation at the fixed point itself. The critical behavior is then governed by the competition between the deterministic gap and empirical fluctuations in a shrinking neighborhood of the boundary. The resulting finite-size scaling exponents depend on the local behavior of the threshold distribution and must be determined separately.

\subsection{Critical finite-size scaling for power-law thresholds}
\label{subsec:powerlaw_fss}

We now apply the empirical-process framework of Section~\ref{subsec:empirical_dynamics} to the continuous cascade transition for power-law threshold distributions, $F(\rho)=\rho^\alpha$, with $\alpha>1$. As shown in Section~\ref{sec:det_power_law}, the thermodynamic transition occurs at $\rho_0^c=1-1/\alpha$, 
where the fully active state $\rho^*=1$ becomes marginal. Because this marginal fixed point sits on the upper boundary of the domain, the Brownian-bridge noise amplitude $B(F(1))=B(1)=0$ vanishes, invalidating the standard interior Gaussian expansion of Section~\ref{subsec:finite_size_general}.

Near complete activation ($\rho \approx 1$), the natural small order parameter is the number of remaining inactive agents,
\begin{equation}
z_t = N(1-\rho_t),
\end{equation}
which counts the deficit from the fully active state. Substituting $\rho_t = 1 - z_t/N$ and $1 - \rho_{0,N} = (N-M_0)/N$ into the exact finite-$N$ map~(\ref{eq:empirical_map}) cancels the non-seed count $N-M_0$ and yields an exact discrete evolution equation for the inactive population,
\begin{equation}
\label{eq:powerlaw_inactive_map}
z_{t+1}
=
\sum_{i=1}^{N-M_0}
\Theta\left(X_i-1+\frac{z_t}{N}\right),
\end{equation}
starting from $z_0 = N - M_0$. Physically, equation~(\ref{eq:powerlaw_inactive_map}) states that the number of agents still inactive at step $t+1$ equals the number of non-seed agents whose quenched thresholds $X_i$ exceed the current active fraction $1 - z_t/N$.

For a fixed integer $z$, each indicator term in equation~(\ref{eq:powerlaw_inactive_map}) is an independent Bernoulli trial that equals unity with tail probability
\begin{equation}
\label{eq:powerlaw_tail}
q_N(z)
=
1-F\left(1-\frac{z}{N}\right)
=
1-\left(1-\frac{z}{N}\right)^\alpha.
\end{equation}
Thus, at any fixed $z$, the right-hand side of equation~(\ref{eq:powerlaw_inactive_map}) is a binomial random variable with parameters $N-M_0$ and $q_N(z)$. Crucially, however, the values at different $z$ are strongly correlated because they are evaluated on the same quenched realization $\{X_i\}$.

To probe the critical boundary layer where $z \ll N$, we parameterize the realized seed fraction as $\rho_{0,N} = \rho_0^c + \epsilon$, so that $N - M_0 = N(1/\alpha - \epsilon)$. Expanding the tail probability~(\ref{eq:powerlaw_tail}) for $z/N \ll 1$ gives
\begin{equation}
q_N(z)
=
\frac{\alpha z}{N}
-
\frac{\alpha(\alpha-1)}{2}
\frac{z^2}{N^2}
+
O\left(\frac{z^3}{N^3}\right).
\end{equation}
Multiplying by $N-M_0$ and subtracting $z$ gives the expected single-step change (deterministic drift) in the inactive count at fixed $z$:
\begin{equation}
\label{eq:powerlaw_mean_drift}
(N-M_0)q_N(z)-z
=
-\alpha\epsilon z
-
\frac{\alpha-1}{2N}z^2
+
O\left(
|\epsilon|\frac{z^2}{N}
+
\frac{z^3}{N^2}
\right).
\end{equation}
The first term, $-\alpha\epsilon z$, represents the linear driving force away from criticality, whereas the second term, $-\frac{\alpha-1}{2N}z^2$, provides the nonlinear saturation that arrests the cascade when $\epsilon < 0$.

Because $q_N(z) \sim \alpha z/N \ll 1$ in the critical boundary layer, inactive agents represent rare tail events, and the binomial count becomes asymptotically Poissonian. Its variance at fixed $z$ is therefore equal to its mean:
\begin{equation}
\label{eq:powerlaw_variance}
(N-M_0)q_N(z)[1-q_N(z)]
=
z[1+o(1)]
\end{equation}
as $\epsilon\to 0$ and $z/N\to 0$. Consequently, the typical amplitude of sample-to-sample counting fluctuations is $\sqrt{z}$ rather than $\sqrt{N}$. This suppression of noise from $\mathcal{O}(\sqrt{N})$ to $\mathcal{O}(\sqrt{z})$ is the direct microscopic manifestation of the vanishing Brownian-bridge variance at the upper boundary.

The critical finite-size exponents follow immediately from a dominant-balance comparison among the three competing terms in equations~(\ref{eq:powerlaw_mean_drift}) and~(\ref{eq:powerlaw_variance}):
\begin{enumerate}[label=(\roman*)]
\item \emph{Critical deficit exponent:} At exact criticality ($\epsilon=0$), the linear drift vanishes and the nonlinear saturation drift $\sim z^2/N$ balances the stochastic counting fluctuations $\sim \sqrt{z}$ when
\begin{equation}
\frac{z^2}{N}\sim\sqrt{z},
\end{equation}
which yields the characteristic inactive population scale $z\sim N^{2/3}$. Therefore,  the mean final inactive fraction at the critical point decays as
\begin{equation}
\label{eq:scaling_deficit}
1-\langle\rho_\infty(N)\rangle
= \frac{\langle z_\infty \rangle}{N}
\sim N^{-1/3}.
\end{equation}

\item \emph{Critical window exponent:} Away from criticality ($\epsilon \neq 0$), the linear detuning drift $\alpha|\epsilon|z$ becomes comparable to the critical drift $z^2/N \sim N^{1/3}$ (and to the noise $\sqrt{z} \sim N^{1/3}$) at $z \sim N^{2/3}$ when $|\epsilon| N^{2/3} \sim N^{1/3}$, which implies
\begin{equation}
|\epsilon| \sim N^{-1/3}.
\end{equation}
\end{enumerate}

\begin{figure}
\centering
\includegraphics[width=\textwidth]{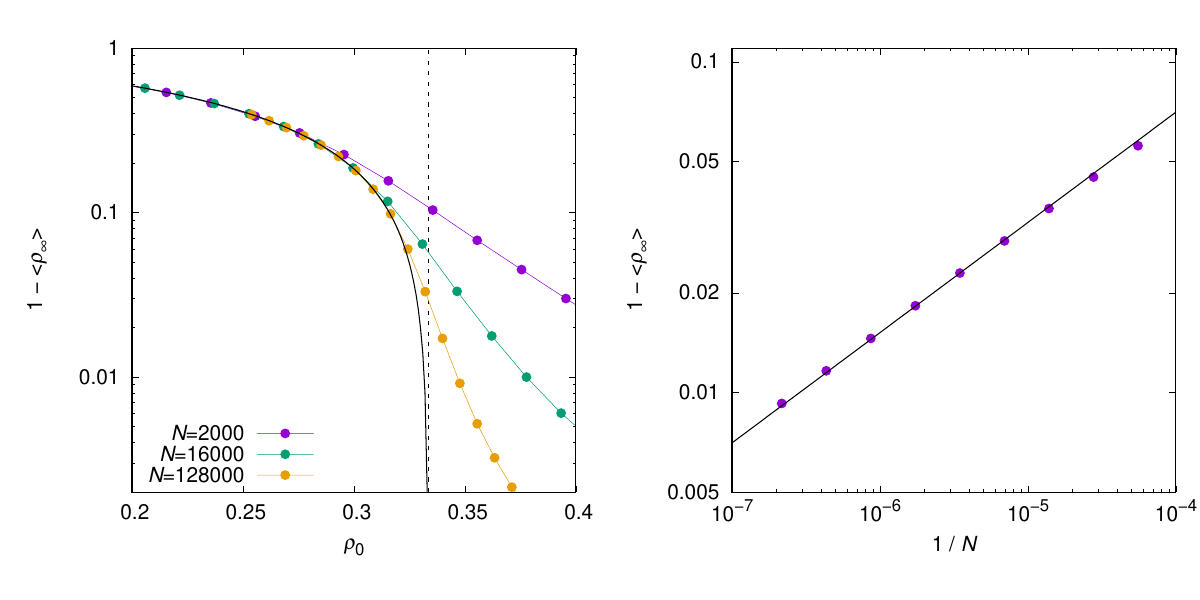}
\caption{
Finite-size scaling of the continuous cascade transition
for power-law thresholds $F(\rho)=\rho^\alpha$ with
$\alpha=1.5$.
Left panel: Mean final inactive fraction
$1-\langle\rho_\infty(N)\rangle$ as a function of
the initial seed density $\rho_0$, for
$N=2000$, $16000$, and $128000$.
The vertical dashed line marks the deterministic
critical density $\rho_0^c=1/3$, and the solid
black curve represents the deterministic prediction.
Right panel: Mean final inactive fraction at
criticality versus inverse population size $1/N$
on logarithmic axes. The Monte Carlo data are
consistent with the predicted $N^{-1/3}$ decay.
The solid line is a fit of the form $aN^{-1/3}$,
with $a=1.52$. Statistical error bars are smaller
than the symbols.
}
\label{fig:power_panels}
\end{figure}

These balances identify the natural finite-size
scaling variables
$u=(\rho_{0,N}-\rho_0^c)N^{1/3}$
and $s=z/N^{2/3}$, suggesting the scaling form
\begin{equation}
\label{eq:universal_data_collapse}
1-\langle\rho_\infty(N)\rangle
\sim N^{-1/3}\Phi_\alpha(u).
\end{equation}
Although an explicit expression for the scaling
function $\Phi_\alpha(u)$ is not needed here,
its asymptotic behavior can be inferred by
matching the thermodynamic result
$1-\rho_\infty \simeq
2\alpha(\rho_0^c-\rho_0)/(\alpha-1)$.
In the subcritical limit $u\to-\infty$,
this matching requires
\begin{equation}
\label{eq:powerlaw_negative_asymptotic}
\Phi_\alpha(u)\sim
-\frac{2\alpha}{\alpha-1}u.
\end{equation}
Conversely, $\Phi_\alpha(u)\to0$ as
$u\to+\infty$, consistently with complete
activation above the critical seed density.
At criticality, the scaling form predicts
$1-\langle\rho_\infty(N)\rangle
\sim \Phi_\alpha(0)N^{-1/3}$,
in agreement with the numerical results
shown in Fig.~\ref{fig:power_panels}.

Figure~\ref{fig:power_panels} tests these
predictions against Monte Carlo simulations
for $\alpha=1.5$, corresponding to
$\rho_0^c=1/3$, using $10^5$ independent
threshold realizations. The left panel
shows the progressive rounding of the
deterministic transition at finite $N$,
with the numerical results approaching
the thermodynamic prediction as the
population size increases. The right
panel confirms the predicted critical
decay
$1-\langle\rho_\infty(N)\rangle\sim N^{-1/3}$,
with a fitted amplitude $a=1.52$.

These results illustrate the distinctive
finite-size behavior of the continuous
cascade transition. Unlike the hybrid
transition governed by an interior
saddle-node, the critical point here
lies at the fully active boundary,
where the suppression of threshold
fluctuations leads to the
$N^{-1/3}$ scaling regime.

\subsection{Finite-size scaling of discontinuous cascade transitions}
\label{subsec:discontinuous_fss}

We now turn to the discontinuous cascade transition associated with the symmetric Beta threshold distribution, with shape parameters $\alpha=\beta=2$ and cumulative distribution
\begin{equation}
F(\rho)=3\rho^2-2\rho^3.
\end{equation}
As established in Section~\ref{sec:det_reg}, the deterministic dynamics exhibits a saddle-node bifurcation at
\begin{equation}
\rho_0^c=\frac{1}{9},
\qquad
\rho^*=\frac{1}{4}.
\end{equation}
Geometrically, for $\rho_0 < \rho_0^c$, the deterministic gap function $H(\rho;\rho_0)$ rises from $-\rho_0 < 0$ to cross zero at a stable low-activity fixed point $\rho_-$, forms a positive hump ($H > 0$) that acts as an activation barrier blocking further growth, and drops back below zero past an unstable threshold $\rho_+$. As $\rho_0$ approaches $\rho_0^c$, the barrier height shrinks to zero and the stable and unstable fixed points coalesce at the marginal bottleneck $\rho^* = 1/4$; for $\rho_0 > \rho_0^c$, the barrier disappears entirely and the cascade proceeds to complete consensus, $\rho_\infty = 1$.

In finite populations, sample-to-sample fluctuations of the empirical threshold distribution smooth this discontinuous jump into a continuous crossover. Near criticality, the deterministic activation barrier is nearly flat, so realizations with the same initial seed fraction can experience opposite macroscopic fates: trajectories whose random gap retains a small positive barrier stall near the bottleneck $\rho^*$, whereas those whose fluctuating gap dips below zero pass freely through the bottleneck and activate the entire population. This coexistence of stalled and fully active realizations generates a bimodal distribution of the final active fraction.

Using the empirical-gap formulation of Section~\ref{subsec:empirical_dynamics}, we analyze these two competing outcomes separately by examining both the probability of crossing the bottleneck and the final active fraction conditioned on trajectories that fail to reach consensus.

\subsubsection{Bottleneck crossing and Gaussian scaling}
\label{subsubsec:bimodal_distribution_crossing}

We define the crossing probability as the probability that a finite-$N$ cascade overcomes the bottleneck and reaches complete consensus,
\begin{equation}
\label{eq:pcross_definition}
P_{\mathrm{cross}}(N,\rho_{0,N})
=
P_N(N;M_0)
=
Q_N^{(M_0)}.
\end{equation}
While the exact recursion~(\ref{eq:Qk_gen}) determines $P_{\mathrm{cross}}$ at any finite $N$, its asymptotic behavior in the critical region follows directly from the local expansion of the empirical gap around the bottleneck.

Recall that the deterministic gap $H(\rho;\rho_0)$ is given by equation~(\ref{eq:deterministic_gap}) and that at 
 the saddle-node $(\rho^*,\rho_0^c) = (1/4, 1/9)$, both the gap and its first derivative vanish, $H(\rho^*;\rho_0^c)=0$ and $\partial H/\partial\rho=0$, while $\partial H/\partial\rho_0 = -[1-F(\rho^*)] = -27/32$ and $\partial^2 H/\partial\rho^2 = -(1-\rho_0^c)f'(\rho^*) = -8/3$. Expanding $H(\rho;\rho_0)$ to leading order around $(\rho^*,\rho_0^c)$ therefore gives
\begin{equation}
\label{eq:gap_expansion}
H(\rho;\rho_0)
\simeq
-\frac{27}{32}(\rho_0-\rho_0^c)
-\frac{4}{3}(\rho-\rho^*)^2,
\end{equation}
where higher-order terms in the seed-density detuning and the displacement from the bottleneck have been omitted. The negative quadratic curvature confirms that $H(\rho;\rho_0^c)$ has a local maximum at $\rho^*=1/4$, while increasing $\rho_0$ above $\rho_0^c$ shifts the curve downward and eliminates the barrier.

In a finite population, the random gap $H_N(\rho)$ fluctuates around $H(\rho;\rho_{0,N})$ according to equation~(\ref{eq:stochastic_gap_expansion}). Because $\rho^*=1/4$ lies  inside the interval $(0,1)$, the Brownian-bridge noise does not vanish at the bottleneck. Evaluating the variance at $F(\rho^*)=5/32$ yields, to leading order,
\begin{equation}
\label{eq:bottleneck_noise_variance}
\operatorname{Var}[H_N(\rho^*)]
=
\frac{1-\rho_0^c}{N}
F(\rho^*)[1-F(\rho^*)]
=
\frac{15}{128N}.
\end{equation}

Balancing the linear detuning term $-27(\rho_{0,N}-\rho_0^c)/32$ in equation~(\ref{eq:gap_expansion}) against the $\mathcal{O}(N^{-1/2})$ standard deviation of the bottleneck noise motivates the critical scaling variable
\begin{equation}
\label{eq:crossing_scaling_variable}
u=(\rho_{0,N}-\rho_0^c)N^{1/2}.
\end{equation}
For fixed $u$ and large $N$, substituting equations~(\ref{eq:gap_expansion}) and~(\ref{eq:bottleneck_noise_variance}) into equation~(\ref{eq:stochastic_gap_expansion}) at $\rho=\rho^*$ gives
\begin{equation}
\label{eq:bottleneck_random_gap}
\sqrt{N}H_N(\rho^*)
\Rightarrow
-\frac{27}{32}u-\sigma Z,
\qquad
\sigma=\sqrt{\frac{15}{128}},
\end{equation}
where $Z$ is a standard normal random variable.

To leading order in the critical scaling limit, the sign of the empirical gap at the deterministic bottleneck determines whether the cascade stalls or reaches complete activation. A negative value of $H_N(\rho^*)$) favors crossing, whereas a positive value indicates the persistence of an activation barrier. At finite $N$, however, local fluctuations can create a positive barrier even when $H_N(\rho^*)<0$, as discussed below. These effects are subleading and do not modify the limiting Gaussian crossing function.  To leading order, the crossing probability therefore converges to the Gaussian cumulative probability
\begin{equation}
P_{\mathrm{cross}}(N,\rho_{0,N})
\longrightarrow
\Psi(u),
\end{equation}
where
\begin{equation}
\label{eq:crossing_scaling_function}
\Psi(u) = \Pr\left( Z>-\frac{27u}{32\sigma} \right) =  \frac12\left[ 1+\operatorname{erf}(1.742842u) \right].
\end{equation}

This result establishes that the critical transition window shrinks as
\begin{equation}
\label{eq:window_discont}
|\rho_{0,N}-\rho_0^c|\sim N^{-1/2}.
\end{equation}
Moreover, at exact criticality ($u=0$), symmetric Gaussian fluctuations yield an asymptotic crossing probability of one half,
\begin{equation}
\label{eq:pcross_half_limit}
\lim_{N\to\infty}
P_{\mathrm{cross}}(N,\rho_0^c)
=
\frac12,
\end{equation}
where $\rho_0^c$ is approached via a sequence of integer seed counts satisfying $\rho_{0,N}-\rho_0^c=o(N^{-1/2})$.

This $50\%$ split has an immediate consequence for the statistics of the final active fraction. At criticality, the half of realizations that stall accumulate near the bottleneck $\rho^*=1/4$, whereas the remaining half reach $\rho_\infty=1$. The probability density of $\rho_\infty$ therefore converges to a bimodal sum of two Dirac delta peaks,
\begin{equation}
\label{eq:bimodal_limit}
P_N(\rho_\infty)
\Rightarrow
\frac12\delta\left(\rho_\infty-\frac14\right)
+
\frac12\delta(\rho_\infty-1),
\end{equation}
so that the ensemble-averaged active fraction approaches
\begin{equation}
\label{eq:mean_rho_limit}
\lim_{N\to\infty}
\langle\rho_\infty(N)\rangle
=
\frac12\left(\frac14+1\right)
=
\frac58.
\end{equation}
Unlike in regular regimes, the order parameter $\rho_\infty$ is not self-averaging at the critical point: its sample-to-sample variance remains  positive ($\operatorname{Var}(\rho_\infty) \to 9/64$) even in the thermodynamic limit.

\subsubsection{Slow convergence to the scaling function}
\label{subsubsec:crossing_corrections}

Although equation~(\ref{eq:crossing_scaling_function}) captures the exact $N \to \infty$ scaling function, finite-size corrections to $P_{\mathrm{cross}}$ decay exceptionally slowly. This slow convergence arises because the single-point approximation~(\ref{eq:bottleneck_random_gap}) tests the sign of the empirical gap only at the deterministic location $\rho^*$, whereas a trajectory actually stalls whenever the \emph{maximum} of the fluctuating gap over a neighborhood of $\rho^*$ remains positive, $\max_{\rho \approx \rho^*} H_N(\rho) > 0$.

Away from the saddle-node, the deterministic gap bends downward quadratically,
\begin{equation}
H(\rho;\rho_0^c)-H(\rho^*;\rho_0^c)
\sim
-(\rho-\rho^*)^2.
\end{equation}
At the same time, because the empirical process converges to a Brownian bridge, its local increments $B(F(\rho^*+\delta\rho)) - B(F(\rho^*))$ behave like standard Brownian motion over short intervals of width $\delta\rho$, scaling as $(\delta\rho)^{1/2}$. Multiplying by the global $N^{-1/2}$ noise prefactor in equation~(\ref{eq:stochastic_gap_expansion}) gives local stochastic gap variations of magnitude
\begin{equation}
\delta H_{\mathrm{noise}}
\sim
N^{-1/2}(\delta\rho)^{1/2}.
\end{equation}

The random peak of $H_N(\rho)$ can wander away from $\rho^*$ only up to the spatial scale $\delta\rho$ where the downward deterministic curvature matches these upward Brownian increments:
\begin{equation}
(\delta\rho)^2
\sim
N^{-1/2}(\delta\rho)^{1/2}
\quad \Longrightarrow \quad
\delta\rho\sim N^{-1/3}.
\end{equation}
Over this $\mathcal{O}(N^{-1/3})$ neighborhood, both the deterministic drop and the local Brownian excursion of the gap have characteristic magnitude
\begin{equation}
\label{eq:local_gap_correction}
\delta H_N\sim (\delta\rho)^2 \sim N^{-2/3}.
\end{equation}

Comparing this local peak excursion $\delta H_N \sim N^{-2/3}$ to the leading single-point Gaussian fluctuation scale $N^{-1/2}$ yields a relative shift in the effective barrier height of order $N^{-2/3}/N^{-1/2} = N^{-1/6}$. 
Because the Gaussian density of $\sqrt{N}H_N(\rho^*)$ is smooth and of order unity near zero, shifting the crossing threshold by $\mathcal{O}(N^{-1/6})$ modifies the crossing probability at the same order:
\begin{equation}
\label{eq:crossing_correction_general}
P_{\mathrm{cross}}(N,\rho_{0,N})
=
\Psi(u)+O(N^{-1/6})
\end{equation}
for fixed $u$ in the critical window.

At exact criticality ($u=0$), this scaling argument predicts
\begin{equation}
\label{eq:pcross_critical_scaling}
\frac12-
P_{\mathrm{cross}}(N,\rho_0^c)
\sim N^{-1/6}.
\end{equation}
Notice that the correction has a definite positive sign: because the local maximum satisfies $\max_{\rho \approx \rho^*} H_N(\rho) \ge H_N(\rho^*)$, upward Brownian excursions near $\rho^*$ can create a positive barrier even when $H_N(\rho^*) < 0$, trapping extra trajectories and holding $P_{\mathrm{cross}}(N,\rho_0^c)$ below $1/2$. 
Because the exponent $1/6$ is unusually small, substantial finite-size offsets from the limiting curve $\Psi(u)$ persist even for populations of millions of agents.

\begin{figure}[t]
\centering
\includegraphics[width=\textwidth]{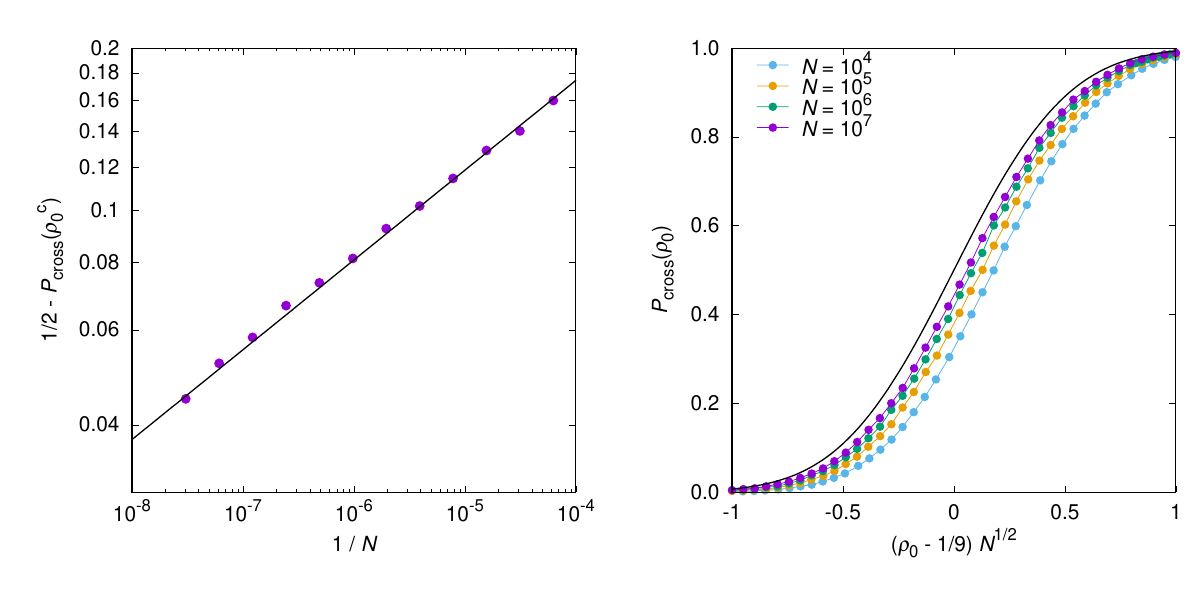}
\caption{
Finite-size scaling of the discontinuous cascade
transition for symmetric Beta thresholds
($\alpha=\beta=2$), with critical seed density
$\rho_0^c=1/9$.
Left panel: Deviation
$1/2-P_{\mathrm{cross}}(N,\rho_0^c)$
versus inverse population size $1/N$
on logarithmic axes. The solid black line
illustrates the predicted $N^{-1/6}$
power-law decay.
Right panel: Crossing probability
$P_{\mathrm{cross}}(N,\rho_0)$ versus the
scaling variable
$u=(\rho_0-\rho_0^c)N^{1/2}$,
for $N=10^4$, $10^5$, $10^6$, and $10^7$.
The solid black curve is the theoretical
asymptotic prediction
$\Psi(u)=[1+\operatorname{erf}(1.742842u)]/2$.
The pronounced separation between the
finite-size curves illustrates the slow
$O(N^{-1/6})$ convergence toward the
limiting scaling function.
Statistical error bars are smaller than
the plotting symbols.
}
\label{fig:discontinuous_panels}
\end{figure}

Figure~\ref{fig:discontinuous_panels} confirms both features against Monte Carlo simulations. The left panel is consistent with  the critical deficit $1/2 - P_{\mathrm{cross}}(N,\rho_0^c)$ decay along the predicted $N^{-1/6}$ power law over several decades in $N$. The right panel plots $P_{\mathrm{cross}}(N,\rho_0)$ against the scaling variable $u=(\rho_0-\rho_0^c)N^{1/2}$ for system sizes from $N=10^4$ to $N=10^7$. While the numerical curves drift systematically toward the analytical Gaussian prediction $\Psi(u)$ as $N$ increases, they remain visibly separated even at $N=10^7$—a direct signature of the slow $\mathcal{O}(N^{-1/6})$ corrections to scaling.

\subsubsection{Conditional active fraction and
saddle-node fluctuations}
\label{subsubsec:conditional_expectation_saddle_node}

To resolve the internal structure of the lower peak in the bimodal distribution~(\ref{eq:bimodal_limit}), we isolate the realizations that fail to reach consensus by defining the conditional stalled expectation
\begin{equation}
\label{eq:cond_expectation_def}
\begin{split}
\langle\rho_\infty\rangle_{\mathrm{stall}}
&=
\mathbb{E}[
\rho_\infty\mid\rho_\infty<1]\\
&=
\frac{1}{1-P_{\mathrm{cross}}(N,\rho_{0,N})}
\sum_{k=M_0}^{N-1}
\frac{k}{N}P_N(k;M_0).
\end{split}
\end{equation}

The finite-size scaling of this conditional observable follows directly from the geometry of the random gap near $\rho^*=1/4$. For a trajectory to stall, the empirical gap must remain positive near the bottleneck, $H_N(\rho^*) \sim N^{-1/2} > 0$. Locally, the random gap forms a downward-opening parabola $H_N(\rho) \simeq H_N(\rho^*) - \frac{4}{3}(\rho-\rho^*)^2$ that intersects zero at two points $\rho^* \pm \sqrt{\frac{3}{4}H_N(\rho^*)}$. Because the monotone cascade starts from $\rho_0 < \rho^*$ and approaches from below, it halts at the lower root $\rho_\infty < \rho^*$, satisfying
\begin{equation}
(\rho^* - \rho_\infty)^2
\sim N^{-1/2}
\quad \Longrightarrow \quad
\rho^* - \rho_\infty
\sim N^{-1/4}.
\end{equation}
(Note that the $\mathcal{O}(N^{-1/3})$ wandering of the parabola's peak derived in Section~\ref{subsubsec:crossing_corrections} is  subleading compared to this $\mathcal{O}(N^{-1/4})$ root displacement.)

When $\rho_0$ is detuned from $\rho_0^c$, balancing the linear detuning term $-27(\rho_{0,N}-\rho_0^c)/32$ in equation~(\ref{eq:gap_expansion}) against the $\mathcal{O}(N^{-1/2})$ bottleneck noise again yields the critical window $|\rho_{0,N}-\rho_0^c| \sim N^{-1/2}$. Combining the $\mathcal{O}(N^{-1/4})$ vertical shift with the $\mathcal{O}(N^{-1/2})$ window width leads to the finite-size scaling ansatz
\begin{equation}
\label{eq:scaling_ansatz_saddle_node}
\rho^*-
\langle\rho_\infty\rangle_{\mathrm{stall}}
=
N^{-1/4}
\Xi\left[
(\rho_0^c-\rho_{0,N})N^{1/2}
\right],
\end{equation}
where $\Xi(v)$ is the scaling function governing the stalled branch.

Far on the subcritical side ($v = (\rho_0^c - \rho_{0,N})N^{1/2} \to +\infty$), fluctuations become negligible and $\langle\rho_\infty\rangle_{\mathrm{stall}}$ must match the deterministic square-root singularity $\rho^* - \rho_\infty \propto (\rho_0^c - \rho_0)^{1/2}$, which imposes the asymptotic power law
\begin{equation}
\Xi(v)\propto v^{1/2},
\qquad v\to+\infty.
\end{equation}
At exact criticality ($v=0$), $\Xi(0)$ is a finite positive constant, yielding the critical scaling law
\begin{equation}
\label{eq:conditional_critical_scaling}
\rho^*-
\langle\rho_\infty\rangle_{\mathrm{stall}}
\sim N^{-1/4}.
\end{equation}

\begin{figure}[htbp]
\centering
\includegraphics[width=\textwidth]{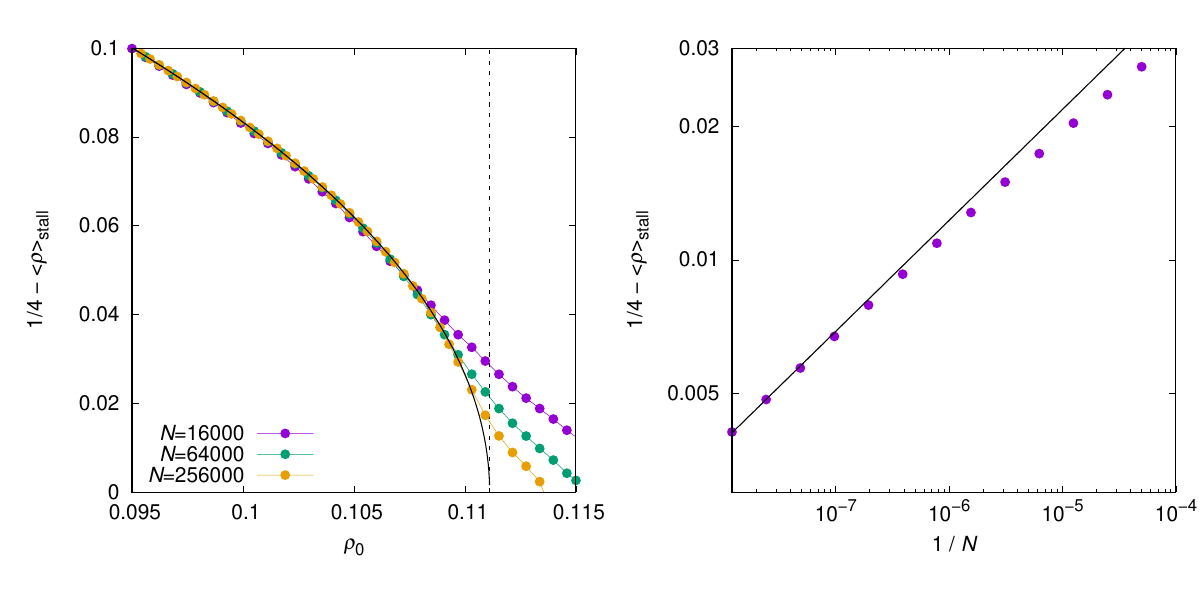}
\caption{
Finite-size regularization of the saddle-node
bottleneck for symmetric Beta thresholds
($\alpha=\beta=2$).
Left panel: Conditional stalled-state gap
$1/4-\langle\rho_\infty\rangle_{\mathrm{stall}}$
as a function of the initial seed density
$\rho_0$, for $N=16000$, $64000$, and $256000$.
The solid black curve represents the
deterministic prediction from
equation~(\ref{eq:sol_alpha2_beta2}),
and the vertical dashed line marks
$\rho_0^c=1/9$.
Right panel: Conditional gap at criticality
versus inverse population size $1/N$
on logarithmic axes, for populations
ranging from $N=10^4$ to
$N=8.192\times10^7$.
The solid black line is a power-law fit  $0.388N^{-1/4}$ with the theoretically predicted exponent fixed at $1/4$.
 Statistical error bars are smaller
than the plotting symbols.
}
\label{fig:saddle_node_scaling}
\end{figure}

Figure~\ref{fig:saddle_node_scaling} illustrates this finite-size regularization of the saddle-node branch. 
The left panel shows how the conditional gap approaches the deterministic low-activity branch as the population size increases, with the most pronounced finite-size deviations concentrated near the critical seed density. The right panel tracks the critical gap $1/4 - \langle\rho_\infty\rangle_{\mathrm{stall}}$ at $\rho_0=\rho_0^c$ up to $N=8.192\times 10^7$, confirming the predicted $N^{-1/4}$ power-law decay with prefactor $0.388$ at large $N$ (once subleading corrections have died out).

Together, Figures~\ref{fig:discontinuous_panels} and~\ref{fig:saddle_node_scaling} show how the two observables probe complementary facets of the same critical bottleneck: $P_{\mathrm{cross}}$ measures the global probability that the fluctuating gap dips below zero to unlock a full cascade (suffering slow $N^{-1/6}$ corrections from local peak excursions), whereas $\langle\rho_\infty\rangle_{\mathrm{stall}}$ measures the $\mathcal{O}(N^{-1/4})$ shift of the lower root when a positive barrier survives.

\subsubsection{Connection with hybrid phase transitions}
\label{subsubsec:hybrid_connection}

The saddle-node cascade transition exemplifies a hybrid (or mixed-order) phase transition, combining a discontinuous macroscopic jump in the order parameter with a critical power-law singularity on the low-activity branch~\cite{Dorogovtsev_2006}.

In the thermodynamic limit, the active fraction jumps discontinuously at $\rho_0^c=1/9$ by an amount
\begin{equation}
\Delta\rho_\infty = 1-\rho^* = \frac{3}{4},
\end{equation}
while on the subcritical side ($\rho_0 \to \rho_0^{c-}$) the order parameter approaches $\rho^*$ with a square-root singularity, $\rho^* - \rho_\infty \propto (\rho_0^c - \rho_0)^{\beta_m}$ with order-parameter exponent $\beta_m = 1/2$. In standard hybrid finite-size scaling notation, the critical window scaling $|\rho_0 - \rho_0^c| \sim N^{-1/\bar{\nu}_m}$ identifies the correlation-volume exponent $\bar{\nu}_m = 2$, while the critical rounding $\rho^* - \langle\rho_\infty\rangle_{\mathrm{stall}} \sim N^{-\mu_m}$ gives $\mu_m = 1/4 = \beta_m/\bar{\nu}_m$.

The empirical-process formulation provides a transparent physical mechanism for these exponents and helps place the present transition within the broader context of hybrid phase transitions. Similar combinations of a discontinuous order-parameter jump and critical singularities have been reported in $k$-core percolation on random graphs~\cite{Dorogovtsev_2006}, interdependent networks~\cite{Baxter_2010}, cascading percolation into an absorbing state~\cite{Lee_2016}, and the continuous-trait Axelrod model~\cite{Reia_2026}. In several of these systems, the hybrid behavior is associated with the disappearance of a stable branch through a saddle-node bifurcation of a mean-field self-consistency equation. However, the microscopic sources of randomness and the resulting finite-size fluctuations differ among models. In the present case, the exponents $\beta_m=1/2$, $\bar{\nu}_m=2$, and $\mu_m=1/4$ follow directly from the competition between the quadratic deterministic gap near the saddle-node and the $N^{-1/2}$ fluctuations of the empirical threshold distribution. Thus, while the model shares the characteristic mean-field phenomenology of hybrid transitions, its finite-size scaling mechanism can be identified explicitly in terms of quenched threshold fluctuations.

On the other hand, the microscopic origin of the critical fluctuations here differs fundamentally from structural percolation models. Rather than arising from topological pruning of a random graph, the fluctuations in our model stem from sample-to-sample variations in the empirical distribution of quenched individual thresholds, described asymptotically by a correlated Brownian bridge. A key consequence of this functional disorder is that while the leading exponents match standard mean-field hybrid values, local Brownian excursions of the empirical gap over an $\mathcal{O}(N^{-1/3})$ neighborhood of the saddle-node generate unusually slow $\mathcal{O}(N^{-1/6})$ corrections to the crossing probability—a feature absent in simple uncorrelated Gaussian approximations.

Furthermore, these scaling laws are not restricted to the symmetric Beta distribution with $\alpha=\beta=2$. They are expected to hold for generic interior saddle-node transitions within the Beta threshold family, provided the deterministic gap has nonvanishing quadratic curvature at the bottleneck, the seed-density detuning enters linearly, and the empirical threshold fluctuations have nonzero variance at the critical state. The choice $\alpha=\beta=2$  allows the critical seed density and bottleneck location to be expressed in closed rational form, making the underlying scaling mechanism particularly transparent.

\section{Discussion}
\label{sec:disc}

In this work, we have generalized the classic Granovetter threshold model from the single-instigator scenario ($M_0=1$)~\cite{Fontanari_2026} to an extensive initial seed of activists ($M_0=\rho_0 N$). From a sociological perspective, this extension provides a framework for quantifying social tipping points, particularly the minimum fraction of initially committed individuals required to overcome collective inertia and trigger widespread adoption. By incorporating a macroscopic seed density, the model captures the role of organized critical masses in social mobilization, with potential applications to collective movements~\cite{Watts_2002,Centola_2018}, opinion shifts~\cite{Galam_2012}, and behavioral contagion~\cite{Centola_2007,Ugander_2012}.

From the standpoint of statistical physics, the introduction of an extensive seed density reveals a rich phase diagram containing both continuous and discontinuous cascade transitions. Their finite-size behavior can be analyzed through the empirical cumulative distribution of quenched individual thresholds, whose fluctuations are governed by classical results from empirical-process theory~\cite{van_der_Vaart_1998}. This formulation provides an exact description of the finite-population dynamics and identifies the stochastic fluctuations responsible for rounding the deterministic transitions. In particular, the continuous transition associated with power-law thresholds exhibits a critical window of width $N^{-1/3}$ and a mean inactive fraction that decays as $N^{-1/3}$ at criticality. By contrast, the discontinuous transition associated with an interior saddle-node has a critical window of width $N^{-1/2}$, with conditional fluctuations of the stalled state scaling as $N^{-1/4}$.

A central result is that the discontinuous cascade transition belongs to the broad family of hybrid, or mixed-order, phase transitions. Such transitions combine a discontinuous change in the order parameter with critical singularities normally associated with continuous transitions. For the symmetric Beta distribution with $\alpha=\beta=2$, the deterministic final active fraction jumps by $\Delta\rho_\infty=3/4$ at the critical seed density $\rho_0^c=1/9$. Nevertheless, the stable low-activity branch approaches the marginal state $\rho^*=1/4$ with the square-root singularity $\rho^*-\rho_\infty\propto (\rho_0^c-\rho_0)^{1/2}$. The corresponding finite-size scaling exponents are $\beta_m=1/2$, $\bar{\nu}_m=2$, and $\mu_m=1/4=\beta_m/\bar{\nu}_m$. Thus, the discontinuity is accompanied by a diverging response to changes in the seed density and anomalously large fluctuations near the marginal bottleneck.

The coexistence of discontinuity and criticality has a long history in statistical physics. An early example is Fisher's treatment of cooperative transitions in biopolymers, where the statistical properties of loops play a central role in determining the character of the denaturation transition~\cite{Fisher_1966}. Another classical example is the one-dimensional Ising model with inverse-square interactions, studied by Thouless~\cite{Thouless_1969}. Subsequent work has established mixed-order behavior in several equilibrium systems, including models involving domain-wall and
loop statistics~\cite{Bar_2014}. These examples demonstrate that a discontinuity in an order parameter need not exclude critical singularities or a diverging correlation length.

The present model shares this general phenomenology but differs from the classical equilibrium examples in its underlying mechanism. The cascade dynamics is irreversible and intrinsically nonequilibrium: there is no Hamiltonian, temperature, or detailed-balance condition governing the evolution. The discontinuous transition does not result from the coexistence of competing thermodynamic free-energy minima. Instead, it arises from the disappearance of a stable low-activity fixed point through a saddle-node bifurcation~\cite{Strogatz_2015}. The critical singularity is therefore a consequence of the local geometry of the deterministic self-consistency equation, rather than of equilibrium fluctuations between thermodynamic phases.

A second important distinction concerns the meaning of critical correlations. In spatial equilibrium models, mixed-order transitions may be accompanied by a diverging correlation length, which characterizes collective fluctuations over increasingly large distances. In the globally coupled Granovetter model, agents respond to the same population-wide active fraction, and there is no underlying spatial metric or independent spatial correlation length. The relevant critical fluctuations instead arise from the empirical distribution of quenched thresholds. Their correlations are described by a Brownian bridge, while the finite-size scaling laws are expressed directly in terms of the population size $N$. Thus, the present transition shares the mixed-order phenomenology of classical models without requiring the same microscopic origin of criticality.

Closer analogies can be found in hybrid transitions associated with network processes, including $k$-core percolation and interdependent networks~\cite{Dorogovtsev_2006,Baxter_2010,Lee_2016}.
In these systems, a discontinuous change in the surviving or functional fraction can coexist with a square-root singularity near a critical point. The similarity reflects the common saddle-node structure of the corresponding mean-field self-consistency equations. Nevertheless, the microscopic dynamics is different. In $k$-core percolation, the transition results from recursive removal of vertices that fail to satisfy a connectivity constraint, whereas in the Granovetter model it results from irreversible activation of individuals whose thresholds are exceeded by the collective active fraction. Moreover, randomness in the present model enters explicitly through independent quenched thresholds, rather than through the topology of a random network. These distinctions affect the stochastic processes that govern finite-size fluctuations and their corrections to asymptotic scaling.

The finite-population behavior of the saddle-node transition reveals another important feature of this hybrid criticality. At the critical seed density, the probability of reaching complete activation approaches $1/2$, while the complementary fraction of realizations stalls near $\rho^*=1/4$. Consequently, the distribution of the final active fraction converges to a bimodal mixture of two macroscopic outcomes, and its variance remains finite as $N\to\infty$. In particular, $\langle\rho_\infty\rangle\to5/8$, rather than the marginal value $1/4$ selected by the deterministic dynamics at exact criticality. This difference reflects the noncommutativity of the thermodynamic and long-time limits at the marginal transition and illustrates how arbitrarily small sample fluctuations can select macroscopically distinct absorbing states.

The empirical-process approach also reveals unusually slow convergence to the asymptotic critical behavior. Although the width of the discontinuous transition scales as $N^{-1/2}$, the crossing probability approaches its limiting Gaussian scaling function with corrections of order $N^{-1/6}$. These corrections originate from local fluctuations of the empirical gap in an $N^{-1/3}$ neighborhood of the saddle-node, where Brownian increments compete with the quadratic deterministic curvature. As demonstrated by our simulations, substantial deviations from the limiting scaling function persist even for populations containing $10^7$ agents. This observation underscores the importance of distinguishing the leading critical exponents from the rate at which their asymptotic scaling forms become observable.

An additional feature of the Granovetter model is that the nature of the transition can be changed by varying the distribution of individual thresholds while keeping the global interaction rule unchanged. Power-law distributions with $\alpha>1$ exhibit a continuous transition controlled by a marginal fixed point at the fully active boundary, where empirical fluctuations are suppressed. In contrast, threshold distributions supporting an interior saddle-node exhibit a hybrid transition with nonvanishing Gaussian fluctuations at the bottleneck. The different finite-size exponents, $1/3$ and $1/2$ for the respective critical windows, arise directly from this distinction between boundary and interior marginality. Thus, the model provides a simple setting in which different critical mechanisms emerge from changes in threshold heterogeneity rather than from changes in interaction topology or dynamical rules.

Beyond their theoretical significance, these results have implications for the interpretation of social tipping points in finite populations. In real-world networks and experimental settings, populations are necessarily finite~\cite{Centola_2018,Bond_2012}, and the outcome of a cascade may remain uncertain even when the initial seed density is close to its deterministic critical value. The finite-size scaling laws quantify how this uncertainty depends on population size and show that, near a hybrid transition, nominally identical populations may exhibit radically different collective outcomes. In particular, the slow convergence of crossing probabilities suggests that deterministic critical thresholds alone may be insufficient to predict cascade success reliably in large but finite populations.

Finally, Granovetter himself identified several natural extensions that would increase the empirical realism of threshold models~\cite{Granovetter_1978}. One important example is heterogeneous
social influence, where active acquaintances exert different pressures depending on the strength of their relationships. A network formulation with weighted ties $w_{ij}$ and binary activity variables $s_j$ would allow an agent $i$ to activate when the weighted active-neighbor fraction $\sum_j w_{ij}s_j/\sum_j w_{ij}$ exceeds its individual threshold~\cite{Kempe_2003}. In this setting, the dynamics depends on the spatial or network arrangement of active individuals, rather than solely on their total number, and the scalar mean-field description generally ceases to be exact~\cite{Watts_2002,Gleeson_2007}. Similarly, allowing competing and mutually exclusive social choices~\cite{Ishikawa_2026} introduces multiple coupled order parameters and additional path-dependent dynamical mechanisms. The globally coupled extensive-seed model studied here provides a tractable reference against which the effects of network heterogeneity, competing adoption processes, and more complex social interactions can be assessed.

\section*{Acknowledgments}

BYSI is supported by Fundação de Amparo à Pesquisa do Estado de São Paulo (FAPESP) grant number 26/06155-6.  JFF is partially supported by  Conselho Nacional de Desenvolvimento Cient\'{\i}fico e Tecnol\'ogico  grant number 300200/2026-9.

\end{document}